\documentclass[english,aps,superscriptaddress,manuscript]{revtex4-1}

\usepackage[T1]{fontenc}
\usepackage[latin9]{inputenc}
\usepackage{textcomp}
\usepackage{amsmath}
\usepackage{amssymb}
\usepackage{graphicx}
\usepackage{bm}
\usepackage{esint}
\usepackage{xcolor}

\usepackage{babel}
\begin{document}

\title{Diffusion of Colloidal Rods in Corrugated Channels}

\author{Xiang Yang}

\affiliation{School of Physics and Astronomy and Institute of Natural Sciences,
Shanghai Jiao Tong University, Shanghai, China}

\author{Qian Zhu}

\affiliation{School of Physics and Astronomy and Institute of Natural Sciences,
Shanghai Jiao Tong University, Shanghai, China}

\author{Chang Liu}

\affiliation{School of Physics and Astronomy and Institute of Natural Sciences,
Shanghai Jiao Tong University, Shanghai, China}

\author{Wei Wang}

\affiliation{School of Material Science and Engineering, Harbin Institute of Technology,
Shenzhen Graduate School, Shenzhen,China}

\author{Yunyun Li}

\affiliation{Center for Phononics and Thermal Energy Science, School of Physics
Science and Engineering, Tongji University, Shanghai, China}

\author{Fabio Marchesoni }

\affiliation{Center for Phononics and Thermal Energy Science, School of Physics
Science and Engineering, Tongji University, Shanghai, China}

\affiliation{Dipartimento di Fisica, Universit\`a di Camerino, I-62032 Camerino,
Italy}

\author{Peter H\"{a}nggi}

\affiliation{Institut f\"{u}r Physik, Universit\"{a}t Augsburg, D-86135 Augsburg, Germany}

\affiliation{Nanosystems Initiative Munich, Schellingstrasse 4, D-80799 M\"{u}nchen,
Germany}

\author{H. P. Zhang}

\affiliation{School of Physics and Astronomy and Institute of Natural Sciences,
Shanghai Jiao Tong University, Shanghai, China}

\affiliation{Collaborative Innovation Center of Advanced Microstructures, Nanjing,
China}

\date{\today}
\begin{abstract}
In many natural and artificial devices  diffusive transport takes place in
confined geometries with corrugated boundaries. Such boundaries cause both
entropic and hydrodynamic effects, which have been studied only for the
case of spherical particles. Here we experimentally investigate diffusion of
particles of elongated shape confined into a corrugated quasi-two-dimensional
channel. Elongated shape causes complex excluded-volume interactions between particle and channel walls which reduce the accessible configuration space and lead to novel entropic free energy effects. The extra rotational degree of freedom also
gives rise to a complex diffusivity matrix that depends on both the particle
location and its orientation. We further show how to extend the standard
Fick-Jacobs theory to incorporate combined hydrodynamic and entropic effects,
so as, for instance, to accurately predict experimentally measured mean first
passage times along the channel. Our approach can be used as a
generic method to describe translational diffusion of anisotropic particles in corrugated channels. {\normalsize \par}
\end{abstract}

\pacs{XXX}


\maketitle

Diffusive transport through micro-structures such as occurring in
porous media \citep{Berkowitz2006,Skaug2015}, micro/nano-fluidic
channels \citep{Kettner2000,Matthias2003,Yang2017b,skaug2018nanofluidic,slanina2016inertial}
and living tissues \citep{Zhou2008,Bressloff2013}, is ubiquitous
and attracts evergrowing attention from physicists \citep{Hanggi2009,Burada2009},
mathematicians \citep{Benichou2014}, engineers \citep{Berkowitz2006},
and biologists \citep{Zhou2008,Hofling2013,Bressloff2013}. A common
feature  of these systems are confining boundaries of irregular shapes. Spatial confinement can fundamentally
change equilibrium and dynamical properties of a system by both limiting
the configuration space accessible to its diffusing components \citep{Hanggi2009}
and increasing the hydrodynamic drag \citep{Deen1987} on them.

An archetypal model to study confinement effects consists of a spherical
particle diffusing in a corrugated narrow channel, which mimics directed
ionic channels \citep{IonChannel}, zeolites \citep{Zeolites}, and
nanopores \citep{Wanunu2010}. In this context, Jacobs \citep{Jacobs1967}
and Zwanzig \citep{Zwanzig1992} proposed a theoretical formulation
to account for the entropic effects stemming from constrained transverse
diffusion. Focusing on the transport (channel) direction, they assumed
that the transverse degrees of freedom (d.o.f's) equilibrate sufficiently
fast and can, therefore, be eliminated adiabatically by means of an
approximate projection scheme. In first order, they derived a reduced
diffusion equation in the channel direction, known as the Fick-Jacobs
(FJ) equation. Numerical investigations \citep{Reguera2001,Kalinay2006,Reguera2006,Berezhkovskii2007,Burada2009}
demonstrated that the FJ equation provides a useful tool to accurately
estimate the entropic effects for confined pointlike particles. However,
our recent experiments \citep{Yang2017b} evidentiated that hydrodynamic
effects for finite size particles cannot be disregarded if the channel
and particle dimensions grow comparable. In order to incorporate such
hydrodynamic corrections, the FJ equation must then be amended in terms of the experimentally measured particle diffusivity.

Previous studies on confined diffusion focused mostly on spherical
particles, for which only the translational d.o.f's were considered. However, particles in practical applications appear inherently more complex in exhibiting anisotropic shape and possessing additional degrees of freedom other than translational. For example, anisotropic particles, such as colloids \citep{Han2006,Chakrabarty2013,kasimov2016diffusion,Hofling2008,SACANNA201196},
artificial and biological filaments \citep{Fakhri2010,Ward2015},
DNA strands \citep{PhysRevLett.94.196101,0953-8984-22-45-454109}
and microswimmers \citep{Bechinger2016,liu2016bimetallic}, exhibit complex coupling between rotation and translation, even in the
absence of geometric constraints. How can complex shape and additional d.o.f's such as rotation alter the current picture of confined diffusion? 
Here, we address this open question and study how a colloidal rod
diffuses in a quasi-two-dimensional (2D) corrugated channel \citep{Wu2015a}. Our experiments
reveal that the interplay of channel's spatial modulation, rod's shape and 
rotational dynamics causes substantial hydrodynamic and entropic
effects. We succeed to extend the standard FJ theory to incorporate
 both effects; the resulting theory accurately predicts
the experimentally measured mean first-passage times (MFPT's) associated
with rod translation along the channel.

\begin{figure}
\includegraphics[width=8cm]{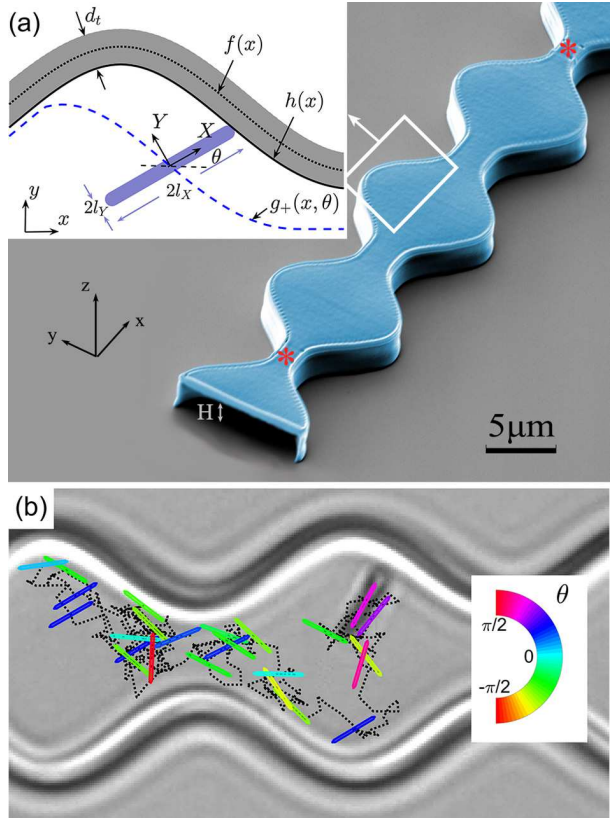}
\caption{(a) Electron scanning image of a thin channel ($H=$ 1.0 $\mu$m, $\alpha={7}/{8}$).
Narrow openings at the two ends are marked by red asterisks.
The inset illustrates a section of the channel
with laser-scanning contour, $f(x)$, wall inner boundary,
$h(x)$, and  upper effective boundary, $g_{+}(x,\theta)$, delimiting
the region accessible to the center of a rod with a given tilting angle,
$\theta$. Rod's length and width and wall thickness are denoted respectively
by $2l_{X}$, $2l_{Y}$ and $d_{t}$. The coordinates $x,y,z$ and $X,Y$ refer respectively
to the laboratory and body frames. (b) Sample
of time discretized trajectory (dotted line) for a rod with $l_{X}=$1.5
$\mu$m in a tall channel ($H=$ 2.0 $\mu$m, $\alpha=1$);
the rod's orientation at different times is also reported according
to the depicted color-code.}
\end{figure}
\textit{Experimental setup.} Our channels were fabricated on a coverslip
by means of a two-photon direct laser writing system, which solidifies
polymers according to a preassigned channel profile, $f(x)$, with
a submicron resolution \citep{Yang2017b}. As depicted in Fig.1 (a), the quasi-2D channel
has a uniform height (denoted by $H$). In the central region, the
periodically curved lateral walls form cells of length $L$ with inner
boundaries a distance $y=\pm h(x)$ away from the channel's axis.
The preassigned profile $f(x)$ is given the form of a cosine, which tapers
off to a constant in correspondence with the cell connecting ducts,
or necks, that
is

\begin{equation}
f(x)=\begin{cases}
\begin{array}{c}
\frac{1}{2}(f_{w}+f_{n})+\frac{1}{2}(f_{w}-f_{n})\mathrm{cos}(\frac{2\pi x}{\alpha L}),|x|<\frac{\alpha L}{2}\\
f_{n},\frac{\alpha L}{2}\leq|x|<\frac{1}{2}L
\end{array} & .\end{cases}
\end{equation} The minimum (maximum) half-width of $f(x)$ is denoted
by $f_{n(w)}$, respectively, whereas $(1-\alpha)L$ is the length
of the neck. Due to the lateral wall thickness $d_{t}=$0.8 $\mu$m
[see inset of Fig. 1(a)],
$f(x)$ and $h(x)$ are separated by a distance $d_t/2$, so that
$f_{n(w)}=h_{n(w)}+ {d_{t}}/{2}$. We changed $f_{n}$ continuously
for fixed $L=$ 12 $\mu$m and $f_{w}=$ 4.6 $\mu$m,
while for the remaining channel parameters we considered two typical
geometries: tall channels ($H=$ 2.0 $\mu$m, $\alpha=1$) and thin
channels ($H=$ 1.0 $\mu$m, $\alpha={7}/{8}$). 

After fabrication, channels were immersed in water with suspended iron-plated gold rods
of width $2l_{Y}=$ 0.3 $\mu$m and length $2l_{X}$, which varies
in the range 1.6-3.2 $\mu$m . Using
a magnet, we dragged a rod into the channel through a narrow entrance, which creates insurmountable entropic barriers to
prevent the rod from exiting the channel. The rod's motion in such
quasi-2D channel was recorded through a microscope at 30 frames per
second for up to 20h \citep{Yang2017b}. We tracked rod trajectories in the imaging plane
and extracted its center coordinates, $(x,y)$, and tilting angle,
$\theta$, by standard particle-tracking
algorithms. We detected {\it no} sizeable rod dynamics in the out-of-plane direction, see Movie S1.mp4 in Supplemental Material \citep{SupportInfo}.

A typical rod trajectory is displayed in Fig. 1(b). The channel boundaries
limit the space accessible to the rod and such a limiting effect depends on the rod's orientation:
the rod  gets closer to the boundary if it is aligned tangent
to the walls. To quantify this orientation dependent effect,
we distributed the recorded rod's center coordinates, $(x,y)$, for
a given orientation, $\theta$, into small bins ($0.26\mu$m$\times0.2\mu$m)
and counted how many times the rod's center was to be found in each
bin. The resulting rod center distributions for three values of $\theta$
are plotted in Fig. 2(a). Nearly uniform distributions demonstrate
that the rod diffuses in a flat energy landscape, whereas sharp drops
of the distributions near the boundaries mark the edge of the
accessible space, consistently with  $y=g_{\pm}(x,\theta)$ computed from the excluded-volume considerations
[see Fig. 2(a)]. The channel boundaries
also affect the rod's orientation. For instance, when the rod is relatively
long, namely for $h_{n}<l_{X}$, then it tends to orient itself parallel
to the channel direction inside the neck region, as illustrated in
the middle panel of Fig. 2(a).

\begin{figure*}
\includegraphics[width=16cm]{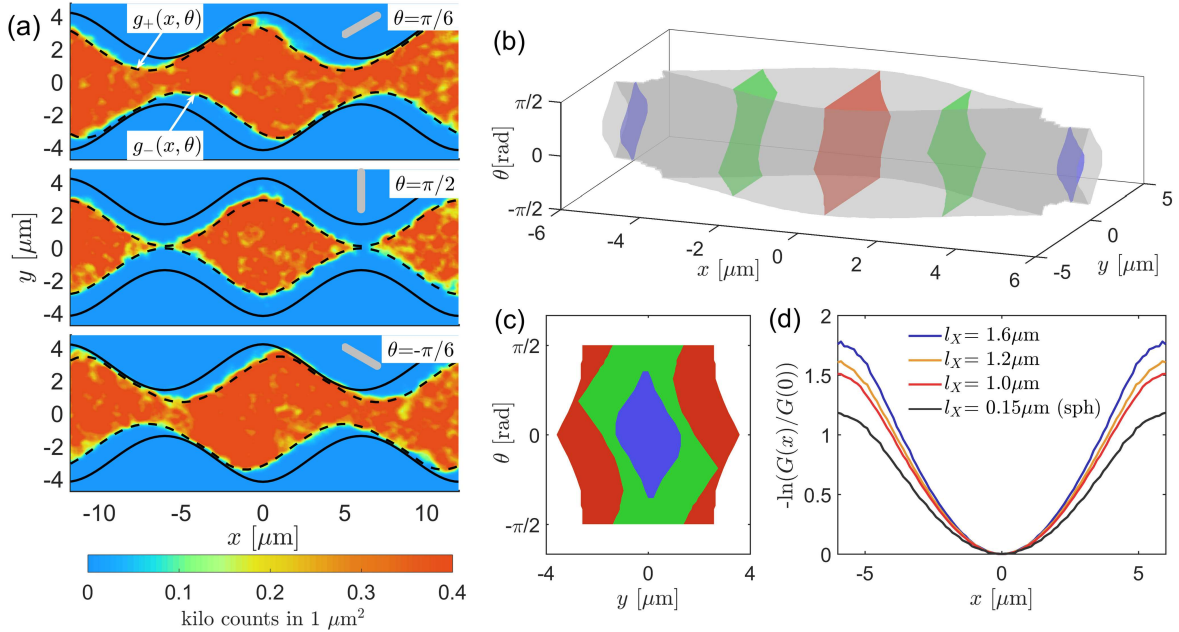}
\caption{(a) Spatial distributions of the rod center for three tilting angles,
$\theta=\frac{\pi}{6}$,$\frac{\pi}{2}$ and $-\frac{\pi}{6}$ . The
channel's inner boundaries, $y=\pm h(x)$, and the tilt-dependent effective
boundaries, $y=g_{\pm}(x,\theta)$, are marked by solid and
dashed lines, respectively. (b) The configuration space accessible
to the confined rod is delimited by the surfaces $y=g_{\pm}(x,\theta)$.
Five cross sections are shown in color; three of them, at $x/L=0,0.22$,
and $0.46$, are displayed in (c). (d) Free-energy profile (in unit
of $k_{B}T$), $-\ln[G(x)/G(0)]$, for different rod lengths (see
text). The black line represents the case of a sphere of radius $l_{X}=0.15$$\mu$m.
Data in (a)-(d) were obtained in a tall channel ($H=$ 2.0 $\mu$m,
$\alpha=1$) with $h_{n}=$1.8 $\mu$m, while the rod used in (a)-(c)
had half-length $l_{X}=1.6$ $\mu$m.}
\end{figure*}

\textit{Fick-Jacobs free-energy.} The rod diffusion can be described
as a random walk in the configuration space ($x,y,\theta$). The dashed
curves $y=g_{\pm}(x,\theta)$ in Fig. 2(a) illustrate how the walls
limit the channel's space
accessible to the rod's center for three different $\theta$ values. From
these curves one can construct a surface in the configuration space,
as shown in Fig. 2(b), and model the motion of the confined rod as
that of a pointlike particle diffusing inside the reconstructed 3D
channel enclosed by that surface. For a rod with length of about 1
$\mu$m,  the relaxation times of $\theta$ and $y$
are short enough for the FJ approach to closely reproduce the long-time diffusion
in the reconstructed 3D channel (see supplemental Sec. IIB \citep{SupportInfo}). To
that end, we integrate the probability density $\rho(x,y,\theta,t)$
to obtain $p(x,t)=\iint\rho(x,y,\theta,t)dyd\theta$ and the corresponding FJ equation
governing its time evolution,

\begin{equation}
\frac{\partial p(x,t)}{\partial t}=\frac{\partial}{\partial x}\left\{ D(x)\left[\frac{\partial p(x,t)}{\partial x}+p(x,t)\frac{\partial}{\partial x}\left(-\ln\frac{G(x)}{G(0)}\right)\right]\right\} .\label{eq:FJequ-1}
\end{equation}
Here, $G(x)=\frac{1}{2\pi}\intop_{-\pi/2}^{\pi/2}[g_{+}(x,\theta)-g_{-}(x,\theta)]d\theta$
represents the area of the $(y,\theta)$ cross section of the reconstructed
3D channel at a given point $x$. Three such cross
sections are plotted in Fig. 2(c). Restrictions in both the center
coordinates, $(x,y)$, and the tilting angle, $\theta$, cause variation
of $G\left(x\right)$. The latter effect is most pronounced in the
neck regions, as illustrated by the blue cross section in Fig. 2(c).
Consequently, the variations of $G\left(x\right)$ modulate the FJ
free-energy profile along the channel. The free-energy potentials
plotted in Fig. 2(d), $-\ln[G(x)/G(0)]$, exhibit
barriers of about 1.8$k_{B}T$ for a rod with a half-length $l_{X}=1.6$ $\mu$m, which is 50\% higher than that of a sphere. This novel entropic effect is induced by particle shape and its strength increases with increasing rod
length.

\textit{Fick-Jacobs effective diffusivity.} Apart from the entropic
potential, the FJ approach introduces an effective
longitudinal diffusivity function, $D(x)$ in Eq. 2. To estimate it, we first
determined the local diffusivity matrix $\mathbb{D}_{IJ}(x,y,\theta)$ of a
rod located at $\left(x,y\right)$ with angle $\theta$, where $I$ and $J$ represent any pair of
coordinates $X$, $Y$ or $\theta$ in the body frame.
As shown in Fig. S1, off-diagonal elements of $\mathbb{D}_{IJ}(x,y,\theta)$ are small and can be
neglected. The remaining three diagonal elements,
$\mathbb{D}_{XX}$, $\mathbb{D}_{YY}$ and $\mathbb{D}_{\theta\theta}$, exhibit a complicated structure inside the channel and
generally have smaller values near channel boundaries, see
Figs. S1(c)-(e). We also numerically computed the hydrodynamic friction
coefficient matrix and then used the fluctuation-dissipation theorem to
numerically estimate the diffusivity matrix. As shown with Fig. S3, numerical
calculations closely reproduce experimental findings. Diffusivity at the channel center can be computed analytically \citep{HappelBook,Tirado1984,Bitter2017,Lisicki2018} and results are in close (5\% difference) agreement with our findings.

We next transformed  $\mathbb{D}_{IJ}(x,y,\theta)$
from the body frame to the laboratory frame and then, in the spirit of the FJ theory, averaged the element
of the resulting diffusivity matrix in the channel's direction, $\mathbb{D}_{xx}$,
over $y$ and $\theta$ to obtain
\begin{equation}
D_{{\rm ave}}(x)=\langle\mathbb{D}_{xx}\rangle_{y,\theta}=\langle\mathbb{D}_{XX}(x,y,\theta)\cos^{2}\theta+\mathbb{D}_{YY}(x,y,\theta)\sin^{2}\theta-\mathbb{D}_{XY}(x,y,\theta)\sin2\theta\rangle_{y,\theta}.\label{eq:Dx-1}
\end{equation}
Figure 3(a) displays the function $D_{{\rm ave}}(x)$ for three different rod
lengths. While for the shortest rod ($l_{X}=$1.0 $\mu$m) $D_{{\rm ave}}(x)$
exhibits minor variability along the channel, for the longest rod
($l_{X}=$1.6 $\mu$m) $D_{{\rm ave}}(x)$ is about 30\% larger in the neck
regions than at the center of the channel cells. This surprising result can be explained by inspecting the
corresponding angular distributions in Fig. 3(b). While around the
center of channel cell the rods can assume any angle, $\theta$, in the necks their
orientation is predominantly constrained around $\theta=0$, more effectively
as the rod length increases. In Eq. (3) for $D_{{\rm ave}}\left(x\right)$,
contributions of $\mathbb{D}_{XX}$ and $\mathbb{D}_{YY}$ are weighted
respectively by $\cos^{2}\theta$ and $\sin^{2}\theta$, implying that for
angular distributions peaked around $\theta=0$ the weight of
$\mathbb{D}_{XX}$ becomes dominant. Moreover, Figs. S1 and S3 confirm that
$\mathbb{D}_{XX}/\mathbb{D}_{YY}\approx2$ in most of configuration space  \citep{Han2009}, so that  $D_{{\rm ave}}(x)$ in the
neck regions is larger for longer rods. In addition to spatial variation, the
hydrodynamic effects also cause a decrease of the local
diffusivity of up to 25\%, as compared to bulk values (see Supplemental Sec. IIA \citep{SupportInfo}).

\begin{figure}
\includegraphics[width=8cm]{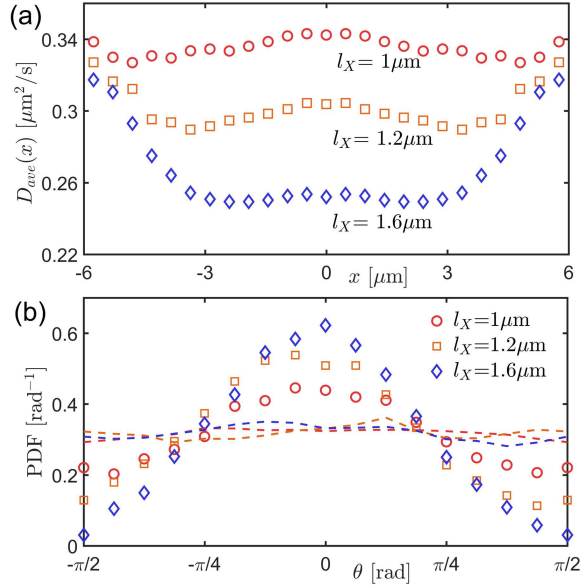}

\caption{(a) Average local diffusivity, $D_{{\rm ave}}(x)$, plotted along
the channel for three rods with $l_{X}$= 1, 1.2 and 1.6 $\mu\textrm{m}$,
see corresponding numerical results in Fig. S3(c). (b) Tilting angle
distributions in the neck regions ( $x=\pm L/2$, symbols),
and at the center of channel cell ($x=0$, dashed lines), for the same  $l_{X}$
as in (a). Data were taken in a tall channel ($H=$ 2.0
$\mu$m, $\alpha=1$) with $h_{n}=$1.4 $\mu$m.}
\end{figure}

We next address the entropic corrections to the local diffusivity,
$D_{{\rm ave}}\left(x\right)$, which in the FJ scheme follow from
the adiabatic elimination of the transverse coordinates \citep{Zwanzig1992,Reguera2001,Berezhkovskii2015}.
Reguera and Rub\'i  proposed heuristic expressions to relate $D(x)$
to $D_{{\rm ave}}(x)$ in narrow 2D and 3D axisymmetric channels \citep{Reguera2001}.
Unfortunately, such expressions do not apply to nonaxisymmetric ``reconstructed''
channels, see Fig. 2(b), where one or more d.o.f.'s are represented by
orientation angles. For this reason we approximated the reconstructed
3D channel of Fig. 2(b) to a quasi-2D channel with half-width $G(x)$, adopted Reguera-Rub\`i  expression \citep{Zwanzig1992,Reguera2001,Berezhkovskii2015} and arrived at the following estimate for $D(x)$,

\begin{equation}
D(x)=\frac{D_{{\rm ave}}(x)}{[1+G'(x)^{2}]^{\frac{1}{3}}}.\label{eq:Deff}
\end{equation}
The validity and corresponding implications of Eq. (4) are discussed
 in Supplemental Sec. IIA \citep{SupportInfo}.

\textit{Mean first-passage times}. With both the entropic potential,
$-\ln G(x)/G(0)$, and the effective logitudinal diffusivity, $D(x)$,
as extracted from the experimental data, one can next apply the FJ equation
to analytically study the diffusive dynamics of confined rods. For
example, we focus on the time duration, $T(\pm\Delta x|0)$,
of the unconditional first passage events that start at $x=0$ and
end up at $x=\pm\Delta x$ [see inset of Fig. 4(a)], regardless of
the fast-relaxing coordinates $y$ and $\theta$. The corresponding
MFPT, $\langle T(\pm\Delta x|0)\rangle$, can then be used to estimate
the asymptotic channel diffusivity in narrow-neck cases, i.e.,
$D_{{\rm ch}}=\lim_{t\rightarrow\infty}\langle[x(t)-x(0)]^{2}\rangle/2t$,
that is $D_{{\rm ch}}=L^{2}/2\langle T(\pm L|0)\rangle$ \citep{Yang2017b}.
Taking advantage of the symmetry properties of the system, Eq. (3) returns an explicit integral expression for the
MFPT \citep{Goel1974,Zwanzig1992}, reading:

\begin{equation}
\langle T_{FJ}(\pm\Delta x|0)\rangle=\intop_{0}^{\Delta x}\frac{d\eta}{G(\eta)D(\eta)}\intop_{0}^{\eta}G(\xi)d\xi.\label{eq:MFPT-1}
\end{equation}
In Fig. 4(a) we compare the predictions of Eq. (5) with the
experimental measurements of $\langle T(\pm\Delta x|0)\rangle$ for
six combinations of $h_{n}$ and $l_{X}$. Without any adjustable
parameters, Eq. (5) yields predictions in excellent agreement
with the experimental data and captures the fast increase of the MFPT
in the neck region. In addition, the validity of our generalized FJ equation has been
systematically explored by extensive
Brownian dynamics simulations in Supplemental Sec. ID \citep{SupportInfo}.

Our experiments were controlled by two geometric parameters: the half-width
of the channel's necks, $h_{n}$, and the rod half-length, $l_{X}$. Numerical and experimental
results in Fig. 4 clearly reveal that the MFPT increases as the ratio
$h_{n}/l_{X}$ decreases. Moreover, provided that the rods are not
too short, $l_{X}>$0.8 $\mu$m, results for different choices of
$h_{n}$ and $l_{X}$, when plotted versus $h_{n}/l_{X}$, collapse
onto a universal curve, as illustrated in Fig. 4(b). This means that,
in the experimental regime investigated here, proportional increases
of $h_{n}$ and $l_{X}$ do not change the MFPT. For a qualitative
explanation of such a property, we notice that increasing $l_{X}$
reduces the available configuration space and, simultaneously, raises
the relevant entropic barriers [Fig. 2(d)]. As a consequence, longer rods, which also
possess smaller diffusivity, $D(x)$ [Fig. 3(a)], tend
to diffuse with longer MFPT's.
On the other hand, increasing $h_{n}$ lowers the entropic barrier,
thus decreasing the MFPT. As quantitatively discussed in Supplemental Sec. IIC \citep{SupportInfo},
these two opposite effects tend to compensate each other, in our experimental regime, as long as the ratio $h_{n}/l_{X}$ is kept constant.

\begin{figure}
\includegraphics[width=8cm]{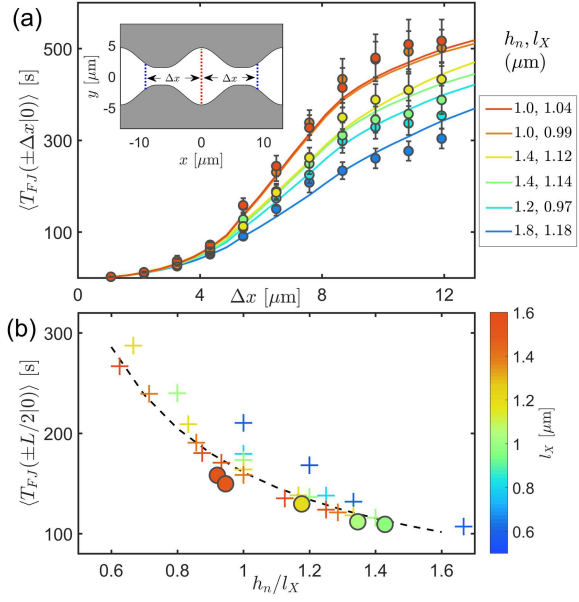}

\caption{(a) MFPT $\langle T(\pm\Delta x|0)\rangle$ vs. $\Delta x$ from experiments (symbols) and theory (curves) in thin channels
($H=$ 1.0 $\mu$m, $\alpha=\frac{7}{8}$) for different values of
the pair ($h_{n}$, $l_{X}$). Inset: vertical dashed segments mark the starting
($x=0$, red) and ending ($x=\pm\Delta x$, blue) positions of the
recorded first passage events. (b) MFPT at $\Delta x=L/2$ vs. $h_{n}/l_{X}$,
measured in tall channels ($H=$ 2.0 $\mu$m, $\alpha=1$) for different
$h_{n}$ and $l_{X}$. Results from experiments and theory are represented
by circles and crosses, respectively; symbols are color-coded according
to the actual value of $l_{X}$. The local diffusivity, $D_{{\rm ave}}\left(x\right)$,
used in the theoretical computations was obtained via finite-element
analysis, see Supplemental Sec. IIA \citep{SupportInfo}.
The dashed line is a guide to the eye.}
\end{figure}

In conclusion, we experimentally measured diffusive transport of colloidal
rods through corrugated planar channels, upon systematically varying
the geometric parameters of the rods and the channel. Anisotropic shape 
significantly impacts particle transport by altering free-energy barriers and particle diffusivity.
Experimental observations
were successfully modeled by generalizing the FJ theory for spherical
particles in terms of an effective longitudinal diffusivity, with hydrodynamic and entropic adjustments, and an FJ free energy including the rotational d.o.f. 

Our method to quantify particle-shape-induced entropic effect (cf. Fig. (2)) is also applicable to model the confined diffusion of even more complex particles, like patchy colloids \citep{SACANNA201196} or polymers
\citep{Fakhri2010,Ward2015}. Such particles possess additional d.o.f.'s,
other than the pure translational ones, and, similarly to the colloidal
rods in our experiments, their description would generally require
higher dimensional configuration spaces. However, as in our work, fast relaxing d.o.f.'s
(``perpendicular'' to the channel direction) may be adiabatically
eliminated and replaced by a reduced free-energy potential [Fig. 2(d)] together with an
effective diffusivity function [Eq. (4) and Fig. 3(a)]. Such a generalization of the FJ approach consequently may serve
as a powerful phenomenological tool to accurately describe the diffusive
transport of real-life particles in directed corrugated narrow channels.

\textit{Acknowledgments - }We acknowledge financial supports of the
NSFC (No. 11774222, 11422427, 11402069) and the Program for Professor of Special
Appointment at Shanghai Institutions of Higher Learning.

\bigskip{}


\begin{thebibliography}{43}%
\makeatletter
\providecommand \@ifxundefined [1]{%
 \@ifx{#1\undefined}
}%
\providecommand \@ifnum [1]{%
 \ifnum #1\expandafter \@firstoftwo
 \else \expandafter \@secondoftwo
 \fi
}%
\providecommand \@ifx [1]{%
 \ifx #1\expandafter \@firstoftwo
 \else \expandafter \@secondoftwo
 \fi
}%
\providecommand \natexlab [1]{#1}%
\providecommand \enquote  [1]{``#1''}%
\providecommand \bibnamefont  [1]{#1}%
\providecommand \bibfnamefont [1]{#1}%
\providecommand \citenamefont [1]{#1}%
\providecommand \href@noop [0]{\@secondoftwo}%
\providecommand \href [0]{\begingroup \@sanitize@url \@href}%
\providecommand \@href[1]{\@@startlink{#1}\@@href}%
\providecommand \@@href[1]{\endgroup#1\@@endlink}%
\providecommand \@sanitize@url [0]{\catcode `\\12\catcode `\$12\catcode
  `\&12\catcode `\#12\catcode `\^12\catcode `\_12\catcode `\%12\relax}%
\providecommand \@@startlink[1]{}%
\providecommand \@@endlink[0]{}%
\providecommand \url  [0]{\begingroup\@sanitize@url \@url }%
\providecommand \@url [1]{\endgroup\@href {#1}{\urlprefix }}%
\providecommand \urlprefix  [0]{URL }%
\providecommand \Eprint [0]{\href }%
\providecommand \doibase [0]{http://dx.doi.org/}%
\providecommand \selectlanguage [0]{\@gobble}%
\providecommand \bibinfo  [0]{\@secondoftwo}%
\providecommand \bibfield  [0]{\@secondoftwo}%
\providecommand \translation [1]{[#1]}%
\providecommand \BibitemOpen [0]{}%
\providecommand \bibitemStop [0]{}%
\providecommand \bibitemNoStop [0]{.\EOS\space}%
\providecommand \EOS [0]{\spacefactor3000\relax}%
\providecommand \BibitemShut  [1]{\csname bibitem#1\endcsname}%
\let\auto@bib@innerbib\@empty
%</preamble>
\bibitem [{\citenamefont {Berkowitz}\ \emph {et~al.}(2006)\citenamefont
  {Berkowitz}, \citenamefont {Cortis}, \citenamefont {Dentz},\ and\
  \citenamefont {Scher}}]{Berkowitz2006}%
  \BibitemOpen
  \bibfield  {author} {\bibinfo {author} {\bibfnamefont {B.}~\bibnamefont
  {Berkowitz}}, \bibinfo {author} {\bibfnamefont {A.}~\bibnamefont {Cortis}},
  \bibinfo {author} {\bibfnamefont {M.}~\bibnamefont {Dentz}}, \ and\ \bibinfo
  {author} {\bibfnamefont {H.}~\bibnamefont {Scher}},\ }\href {\doibase
  10.1029/2005RG000178} {\bibfield  {journal} {\bibinfo  {journal} {Rev.
  Geophys.}\ }\textbf {\bibinfo {volume} {44}},\ \bibinfo {pages} {RG2003}
  (\bibinfo {year} {2006})}\BibitemShut {NoStop}%
\bibitem [{\citenamefont {Skaug}\ \emph {et~al.}(2015)\citenamefont {Skaug},
  \citenamefont {Wang}, \citenamefont {Ding},\ and\ \citenamefont
  {Schwartz}}]{Skaug2015}%
  \BibitemOpen
  \bibfield  {author} {\bibinfo {author} {\bibfnamefont {M.~J.}\ \bibnamefont
  {Skaug}}, \bibinfo {author} {\bibfnamefont {L.}~\bibnamefont {Wang}},
  \bibinfo {author} {\bibfnamefont {Y.~F.}\ \bibnamefont {Ding}}, \ and\
  \bibinfo {author} {\bibfnamefont {D.~K.}\ \bibnamefont {Schwartz}},\ }\href
  {\doibase 10.1021/acsnano.5b00019} {\bibfield  {journal} {\bibinfo  {journal}
  {ACS Nano}\ }\textbf {\bibinfo {volume} {9}},\ \bibinfo {pages} {2148}
  (\bibinfo {year} {2015})}\BibitemShut {NoStop}%
\bibitem [{\citenamefont {Kettner}\ \emph {et~al.}(2000)\citenamefont
  {Kettner}, \citenamefont {Reimann}, \citenamefont {H{\"a}nggi},\ and\
  \citenamefont {Muller}}]{Kettner2000}%
  \BibitemOpen
  \bibfield  {author} {\bibinfo {author} {\bibfnamefont {C.}~\bibnamefont
  {Kettner}}, \bibinfo {author} {\bibfnamefont {P.}~\bibnamefont {Reimann}},
  \bibinfo {author} {\bibfnamefont {P.}~\bibnamefont {H{\"a}nggi}}, \ and\
  \bibinfo {author} {\bibfnamefont {F.}~\bibnamefont {Muller}},\ }\href
  {\doibase 10.1103/PhysRevE.61.312} {\bibfield  {journal} {\bibinfo  {journal}
  {Phys. Rev. E}\ }\textbf {\bibinfo {volume} {61}},\ \bibinfo {pages} {312}
  (\bibinfo {year} {2000})}\BibitemShut {NoStop}%
\bibitem [{\citenamefont {Matthias}\ and\ \citenamefont
  {M{\"u}ller}(2003)}]{Matthias2003}%
  \BibitemOpen
  \bibfield  {author} {\bibinfo {author} {\bibfnamefont {S.}~\bibnamefont
  {Matthias}}\ and\ \bibinfo {author} {\bibfnamefont {F.}~\bibnamefont
  {M{\"u}ller}},\ }\href {\doibase 10.1038/nature01736} {\bibfield  {journal}
  {\bibinfo  {journal} {Nature}\ }\textbf {\bibinfo {volume} {424}},\ \bibinfo
  {pages} {53} (\bibinfo {year} {2003})}\BibitemShut {NoStop}%
\bibitem [{\citenamefont {Yang}\ \emph {et~al.}(2017)\citenamefont {Yang},
  \citenamefont {Liu}, \citenamefont {Li}, \citenamefont {Marchesoni},
  \citenamefont {H{\"a}nggi},\ and\ \citenamefont {Zhang}}]{Yang2017b}%
  \BibitemOpen
  \bibfield  {author} {\bibinfo {author} {\bibfnamefont {X.}~\bibnamefont
  {Yang}}, \bibinfo {author} {\bibfnamefont {C.}~\bibnamefont {Liu}}, \bibinfo
  {author} {\bibfnamefont {Y.}~\bibnamefont {Li}}, \bibinfo {author}
  {\bibfnamefont {F.}~\bibnamefont {Marchesoni}}, \bibinfo {author}
  {\bibfnamefont {P.}~\bibnamefont {H{\"a}nggi}}, \ and\ \bibinfo {author}
  {\bibfnamefont {H.~P.}\ \bibnamefont {Zhang}},\ }\href
  {http://www.pnas.org/content/114/36/9564.abstract} {\bibfield  {journal}
  {\bibinfo  {journal} {Proceedings of the National Academy of Sciences}\
  }\textbf {\bibinfo {volume} {114}},\ \bibinfo {pages} {9564} (\bibinfo {year}
  {2017})}\BibitemShut {NoStop}%
\bibitem [{\citenamefont {Skaug}\ \emph {et~al.}(2018)\citenamefont {Skaug},
  \citenamefont {Schwemmer}, \citenamefont {Fringes}, \citenamefont
  {Rawlings},\ and\ \citenamefont {Knoll}}]{skaug2018nanofluidic}%
  \BibitemOpen
  \bibfield  {author} {\bibinfo {author} {\bibfnamefont {M.~J.}\ \bibnamefont
  {Skaug}}, \bibinfo {author} {\bibfnamefont {C.}~\bibnamefont {Schwemmer}},
  \bibinfo {author} {\bibfnamefont {S.}~\bibnamefont {Fringes}}, \bibinfo
  {author} {\bibfnamefont {C.~D.}\ \bibnamefont {Rawlings}}, \ and\ \bibinfo
  {author} {\bibfnamefont {A.~W.}\ \bibnamefont {Knoll}},\ }\href@noop {}
  {\bibfield  {journal} {\bibinfo  {journal} {Science}\ }\textbf {\bibinfo
  {volume} {359}},\ \bibinfo {pages} {1505} (\bibinfo {year}
  {2018})}\BibitemShut {NoStop}%
\bibitem [{\citenamefont {Slanina}(2016)}]{slanina2016inertial}%
  \BibitemOpen
  \bibfield  {author} {\bibinfo {author} {\bibfnamefont {F.}~\bibnamefont
  {Slanina}},\ }\href@noop {} {\bibfield  {journal} {\bibinfo  {journal} {Phys.
  Rev. E}\ }\textbf {\bibinfo {volume} {94}},\ \bibinfo {pages} {042610}
  (\bibinfo {year} {2016})}\BibitemShut {NoStop}%
\bibitem [{\citenamefont {Zhou}\ \emph {et~al.}(2008)\citenamefont {Zhou},
  \citenamefont {Rivas},\ and\ \citenamefont {Minton}}]{Zhou2008}%
  \BibitemOpen
  \bibfield  {author} {\bibinfo {author} {\bibfnamefont {H.-X.}\ \bibnamefont
  {Zhou}}, \bibinfo {author} {\bibfnamefont {G.}~\bibnamefont {Rivas}}, \ and\
  \bibinfo {author} {\bibfnamefont {A.~P.}\ \bibnamefont {Minton}},\ }\href
  {\doibase 10.1146/annurev.biophys.37.032807.125817} {\bibfield  {journal}
  {\bibinfo  {journal} {Annu. Rev. Biophys.}\ }\textbf {\bibinfo {volume}
  {37}},\ \bibinfo {pages} {375} (\bibinfo {year} {2008})}\BibitemShut
  {NoStop}%
\bibitem [{\citenamefont {Bressloff}\ and\ \citenamefont
  {Newby}(2013)}]{Bressloff2013}%
  \BibitemOpen
  \bibfield  {author} {\bibinfo {author} {\bibfnamefont {P.~C.}\ \bibnamefont
  {Bressloff}}\ and\ \bibinfo {author} {\bibfnamefont {J.~M.}\ \bibnamefont
  {Newby}},\ }\href {\doibase 10.1103/RevModPhys.85.135} {\bibfield  {journal}
  {\bibinfo  {journal} {Rev. Mod. Phys.}\ }\textbf {\bibinfo {volume} {85}},\
  \bibinfo {pages} {135} (\bibinfo {year} {2013})}\BibitemShut {NoStop}%
\bibitem [{\citenamefont {H{\"a}nggi}\ and\ \citenamefont
  {Marchesoni}(2009)}]{Hanggi2009}%
  \BibitemOpen
  \bibfield  {author} {\bibinfo {author} {\bibfnamefont {P.}~\bibnamefont
  {H{\"a}nggi}}\ and\ \bibinfo {author} {\bibfnamefont {F.}~\bibnamefont
  {Marchesoni}},\ }\href {\doibase 10.1103/RevModPhys.81.387} {\bibfield
  {journal} {\bibinfo  {journal} {Rev. Mod. Phys.}\ }\textbf {\bibinfo {volume}
  {81}},\ \bibinfo {pages} {387} (\bibinfo {year} {2009})}\BibitemShut
  {NoStop}%
\bibitem [{\citenamefont {Burada}\ \emph {et~al.}(2009)\citenamefont {Burada},
  \citenamefont {H{\"a}nggi}, \citenamefont {Marchesoni}, \citenamefont
  {Schmid},\ and\ \citenamefont {Talkner}}]{Burada2009}%
  \BibitemOpen
  \bibfield  {author} {\bibinfo {author} {\bibfnamefont {P.~S.}\ \bibnamefont
  {Burada}}, \bibinfo {author} {\bibfnamefont {P.}~\bibnamefont {H{\"a}nggi}},
  \bibinfo {author} {\bibfnamefont {F.}~\bibnamefont {Marchesoni}}, \bibinfo
  {author} {\bibfnamefont {G.}~\bibnamefont {Schmid}}, \ and\ \bibinfo {author}
  {\bibfnamefont {P.}~\bibnamefont {Talkner}},\ }\href {\doibase
  10.1002/cphc.200800526} {\bibfield  {journal} {\bibinfo  {journal}
  {ChemPhysChem}\ }\textbf {\bibinfo {volume} {10}},\ \bibinfo {pages} {45}
  (\bibinfo {year} {2009})}\BibitemShut {NoStop}%
\bibitem [{\citenamefont {Benichou}\ and\ \citenamefont
  {Voituriez}(2014)}]{Benichou2014}%
  \BibitemOpen
  \bibfield  {author} {\bibinfo {author} {\bibfnamefont {O.}~\bibnamefont
  {Benichou}}\ and\ \bibinfo {author} {\bibfnamefont {R.}~\bibnamefont
  {Voituriez}},\ }\href {\doibase 10.1016/j.physrep.2014.02.003} {\bibfield
  {journal} {\bibinfo  {journal} {Phys. Rep.}\ }\textbf {\bibinfo {volume}
  {539}},\ \bibinfo {pages} {225} (\bibinfo {year} {2014})}\BibitemShut
  {NoStop}%
\bibitem [{\citenamefont {Hofling}\ and\ \citenamefont
  {Franosch}(2013)}]{Hofling2013}%
  \BibitemOpen
  \bibfield  {author} {\bibinfo {author} {\bibfnamefont {F.}~\bibnamefont
  {Hofling}}\ and\ \bibinfo {author} {\bibfnamefont {T.}~\bibnamefont
  {Franosch}},\ }\href {\doibase 10.1088/0034-4885/76/4/046602} {\bibfield
  {journal} {\bibinfo  {journal} {Rep. Prog. Phys.}\ }\textbf {\bibinfo
  {volume} {76}},\ \bibinfo {pages} {046602} (\bibinfo {year}
  {2013})}\BibitemShut {NoStop}%
\bibitem [{\citenamefont {Deen}(1987)}]{Deen1987}%
  \BibitemOpen
  \bibfield  {author} {\bibinfo {author} {\bibfnamefont {W.~M.}\ \bibnamefont
  {Deen}},\ }\href {http://dx.doi.org/10.1002/aic.690330902} {\bibfield
  {journal} {\bibinfo  {journal} {AIChE J.}\ }\textbf {\bibinfo {volume}
  {33}},\ \bibinfo {pages} {1409} (\bibinfo {year} {1987})}\BibitemShut
  {NoStop}%
\bibitem [{\citenamefont {Hille}(2001)}]{IonChannel}%
  \BibitemOpen
  \bibfield  {author} {\bibinfo {author} {\bibfnamefont {B.}~\bibnamefont
  {Hille}},\ }\href@noop {} {\emph {\bibinfo {title} {Ion Channels of Excitable
  Membranes}}}\ (\bibinfo  {publisher} {Sinauer Associates},\ \bibinfo {year}
  {2001})\BibitemShut {NoStop}%
\bibitem [{\citenamefont {K{\"a}rger}\ and\ \citenamefont
  {Ruthven}(1992)}]{Zeolites}%
  \BibitemOpen
  \bibfield  {author} {\bibinfo {author} {\bibfnamefont {J.}~\bibnamefont
  {K{\"a}rger}}\ and\ \bibinfo {author} {\bibfnamefont {D.~M.}\ \bibnamefont
  {Ruthven}},\ }\href@noop {} {\emph {\bibinfo {title} {Diffusion in Zeolites
  and Other Microporous Solids}}}\ (\bibinfo  {publisher} {John Wiley, New
  York},\ \bibinfo {year} {1992})\BibitemShut {NoStop}%
\bibitem [{\citenamefont {Wanunu}\ \emph {et~al.}(2010)\citenamefont {Wanunu},
  \citenamefont {Dadosh}, \citenamefont {Ray}, \citenamefont {Jin},
  \citenamefont {McReynolds},\ and\ \citenamefont {Drndic}}]{Wanunu2010}%
  \BibitemOpen
  \bibfield  {author} {\bibinfo {author} {\bibfnamefont {M.}~\bibnamefont
  {Wanunu}}, \bibinfo {author} {\bibfnamefont {T.}~\bibnamefont {Dadosh}},
  \bibinfo {author} {\bibfnamefont {V.}~\bibnamefont {Ray}}, \bibinfo {author}
  {\bibfnamefont {J.~M.}\ \bibnamefont {Jin}}, \bibinfo {author} {\bibfnamefont
  {L.}~\bibnamefont {McReynolds}}, \ and\ \bibinfo {author} {\bibfnamefont
  {M.}~\bibnamefont {Drndic}},\ }\href {\doibase 10.1038/NNANO.2010.202}
  {\bibfield  {journal} {\bibinfo  {journal} {Nat. Nanotechnol.}\ }\textbf
  {\bibinfo {volume} {5}},\ \bibinfo {pages} {807} (\bibinfo {year}
  {2010})}\BibitemShut {NoStop}%
\bibitem [{\citenamefont {Jacobs}(1967)}]{Jacobs1967}%
  \BibitemOpen
  \bibfield  {author} {\bibinfo {author} {\bibfnamefont {M.}~\bibnamefont
  {Jacobs}},\ }\href@noop {} {\emph {\bibinfo {title} {Diffusion Processes}}}\
  (\bibinfo  {publisher} {Springer, New York},\ \bibinfo {year}
  {1967})\BibitemShut {NoStop}%
\bibitem [{\citenamefont {Zwanzig}(1992)}]{Zwanzig1992}%
  \BibitemOpen
  \bibfield  {author} {\bibinfo {author} {\bibfnamefont {R.}~\bibnamefont
  {Zwanzig}},\ }\bibfield  {booktitle} {\emph {\bibinfo {booktitle} {The
  Journal of Physical Chemistry}},\ }\href {\doibase 10.1021/j100189a004}
  {\bibfield  {journal} {\bibinfo  {journal} {J. Phys. Chem.}\ }\textbf
  {\bibinfo {volume} {96}},\ \bibinfo {pages} {3926} (\bibinfo {year}
  {1992})}\BibitemShut {NoStop}%
\bibitem [{\citenamefont {Reguera}\ and\ \citenamefont
  {Rub{\'i}}(2001)}]{Reguera2001}%
  \BibitemOpen
  \bibfield  {author} {\bibinfo {author} {\bibfnamefont {D.}~\bibnamefont
  {Reguera}}\ and\ \bibinfo {author} {\bibfnamefont {J.~M.}\ \bibnamefont
  {Rub{\'i}}},\ }\href {\doibase 10.1103/PhysRevE.64.061106} {\bibfield
  {journal} {\bibinfo  {journal} {Phys. Rev. E}\ }\textbf {\bibinfo {volume}
  {64}},\ \bibinfo {pages} {061106} (\bibinfo {year} {2001})}\BibitemShut
  {NoStop}%
\bibitem [{\citenamefont {Kalinay}\ and\ \citenamefont
  {Percus}(2006)}]{Kalinay2006}%
  \BibitemOpen
  \bibfield  {author} {\bibinfo {author} {\bibfnamefont {P.}~\bibnamefont
  {Kalinay}}\ and\ \bibinfo {author} {\bibfnamefont {J.~K.}\ \bibnamefont
  {Percus}},\ }\href {\doibase 10.1103/PhysRevE.74.041203} {\bibfield
  {journal} {\bibinfo  {journal} {Phys. Rev. E}\ }\textbf {\bibinfo {volume}
  {74}},\ \bibinfo {pages} {041203} (\bibinfo {year} {2006})}\BibitemShut
  {NoStop}%
\bibitem [{\citenamefont {Reguera}\ \emph {et~al.}(2006)\citenamefont
  {Reguera}, \citenamefont {Schmid}, \citenamefont {Burada}, \citenamefont
  {Rub{\'i}}, \citenamefont {Reimann},\ and\ \citenamefont
  {H{\"a}nggi}}]{Reguera2006}%
  \BibitemOpen
  \bibfield  {author} {\bibinfo {author} {\bibfnamefont {D.}~\bibnamefont
  {Reguera}}, \bibinfo {author} {\bibfnamefont {G.}~\bibnamefont {Schmid}},
  \bibinfo {author} {\bibfnamefont {P.~S.}\ \bibnamefont {Burada}}, \bibinfo
  {author} {\bibfnamefont {J.~M.}\ \bibnamefont {Rub{\'i}}}, \bibinfo {author}
  {\bibfnamefont {P.}~\bibnamefont {Reimann}}, \ and\ \bibinfo {author}
  {\bibfnamefont {P.}~\bibnamefont {H{\"a}nggi}},\ }\href {\doibase
  10.1103/PhysRevLett.96.130603} {\bibfield  {journal} {\bibinfo  {journal}
  {Phys. Rev. Lett.}\ }\textbf {\bibinfo {volume} {96}},\ \bibinfo {pages}
  {130603} (\bibinfo {year} {2006})}\BibitemShut {NoStop}%
\bibitem [{\citenamefont {Berezhkovskii}\ \emph {et~al.}(2007)\citenamefont
  {Berezhkovskii}, \citenamefont {Pustovoit},\ and\ \citenamefont
  {Bezrukov}}]{Berezhkovskii2007}%
  \BibitemOpen
  \bibfield  {author} {\bibinfo {author} {\bibfnamefont {A.~M.}\ \bibnamefont
  {Berezhkovskii}}, \bibinfo {author} {\bibfnamefont {M.~A.}\ \bibnamefont
  {Pustovoit}}, \ and\ \bibinfo {author} {\bibfnamefont {S.~M.}\ \bibnamefont
  {Bezrukov}},\ }\href {\doibase 10.1063/1.2719193} {\bibfield  {journal}
  {\bibinfo  {journal} {J. Chem. Phys.}\ }\textbf {\bibinfo {volume} {126}},\
  \bibinfo {pages} {134706} (\bibinfo {year} {2007})}\BibitemShut {NoStop}%
\bibitem [{\citenamefont {Han}\ \emph {et~al.}(2006)\citenamefont {Han},
  \citenamefont {Alsayed}, \citenamefont {Nobili}, \citenamefont {Zhang},
  \citenamefont {Lubensky},\ and\ \citenamefont {Yodh}}]{Han2006}%
  \BibitemOpen
  \bibfield  {author} {\bibinfo {author} {\bibfnamefont {Y.}~\bibnamefont
  {Han}}, \bibinfo {author} {\bibfnamefont {A.~M.}\ \bibnamefont {Alsayed}},
  \bibinfo {author} {\bibfnamefont {M.}~\bibnamefont {Nobili}}, \bibinfo
  {author} {\bibfnamefont {J.}~\bibnamefont {Zhang}}, \bibinfo {author}
  {\bibfnamefont {T.}~\bibnamefont {Lubensky}}, \ and\ \bibinfo {author}
  {\bibfnamefont {A.}~\bibnamefont {Yodh}},\ }\href {\doibase
  10.1126/science.1130146} {\bibfield  {journal} {\bibinfo  {journal}
  {Science}\ }\textbf {\bibinfo {volume} {314}},\ \bibinfo {pages} {626}
  (\bibinfo {year} {2006})}\BibitemShut {NoStop}%
\bibitem [{\citenamefont {Chakrabarty}\ \emph {et~al.}(2013)\citenamefont
  {Chakrabarty}, \citenamefont {Konya}, \citenamefont {Wang}, \citenamefont
  {Selinger}, \citenamefont {Sun},\ and\ \citenamefont
  {Wei}}]{Chakrabarty2013}%
  \BibitemOpen
  \bibfield  {author} {\bibinfo {author} {\bibfnamefont {A.}~\bibnamefont
  {Chakrabarty}}, \bibinfo {author} {\bibfnamefont {A.}~\bibnamefont {Konya}},
  \bibinfo {author} {\bibfnamefont {F.}~\bibnamefont {Wang}}, \bibinfo {author}
  {\bibfnamefont {J.~V.}\ \bibnamefont {Selinger}}, \bibinfo {author}
  {\bibfnamefont {K.}~\bibnamefont {Sun}}, \ and\ \bibinfo {author}
  {\bibfnamefont {Q.~H.}\ \bibnamefont {Wei}},\ }\href {\doibase
  10.1103/PhysRevLett.111.160603} {\bibfield  {journal} {\bibinfo  {journal}
  {Phys. Rev. Lett.}\ }\textbf {\bibinfo {volume} {111}},\ \bibinfo {pages}
  {160603} (\bibinfo {year} {2013})}\BibitemShut {NoStop}%
\bibitem [{\citenamefont {Kasimov}\ \emph {et~al.}(2016)\citenamefont
  {Kasimov}, \citenamefont {Admon},\ and\ \citenamefont
  {Roichman}}]{kasimov2016diffusion}%
  \BibitemOpen
  \bibfield  {author} {\bibinfo {author} {\bibfnamefont {D.}~\bibnamefont
  {Kasimov}}, \bibinfo {author} {\bibfnamefont {T.}~\bibnamefont {Admon}}, \
  and\ \bibinfo {author} {\bibfnamefont {Y.}~\bibnamefont {Roichman}},\
  }\href@noop {} {\bibfield  {journal} {\bibinfo  {journal} {Phys. Rev. E}\
  }\textbf {\bibinfo {volume} {93}},\ \bibinfo {pages} {050602} (\bibinfo
  {year} {2016})}\BibitemShut {NoStop}%
\bibitem [{\citenamefont {Hofling}\ \emph {et~al.}(2008)\citenamefont
  {Hofling}, \citenamefont {Frey},\ and\ \citenamefont
  {Franosch}}]{Hofling2008}%
  \BibitemOpen
  \bibfield  {author} {\bibinfo {author} {\bibfnamefont {F.}~\bibnamefont
  {Hofling}}, \bibinfo {author} {\bibfnamefont {E.}~\bibnamefont {Frey}}, \
  and\ \bibinfo {author} {\bibfnamefont {T.}~\bibnamefont {Franosch}},\ }\href
  {\doibase 10.1103/PhysRevLett.101.120605} {\bibfield  {journal} {\bibinfo
  {journal} {Phys. Rev. Lett.}\ }\textbf {\bibinfo {volume} {101}},\ \bibinfo
  {pages} {120605} (\bibinfo {year} {2008})}\BibitemShut {NoStop}%
\bibitem [{\citenamefont {Sacanna}\ and\ \citenamefont
  {Pine}(2011)}]{SACANNA201196}%
  \BibitemOpen
  \bibfield  {author} {\bibinfo {author} {\bibfnamefont {S.}~\bibnamefont
  {Sacanna}}\ and\ \bibinfo {author} {\bibfnamefont {D.~J.}\ \bibnamefont
  {Pine}},\ }\href {\doibase https://doi.org/10.1016/j.cocis.2011.01.003}
  {\bibfield  {journal} {\bibinfo  {journal} {Current Opinion in Colloid \&
  Interface Science}\ }\textbf {\bibinfo {volume} {16}},\ \bibinfo {pages} {96
  } (\bibinfo {year} {2011})}\BibitemShut {NoStop}%
\bibitem [{\citenamefont {Fakhri}\ \emph {et~al.}(2010)\citenamefont {Fakhri},
  \citenamefont {MacKintosh}, \citenamefont {Lounis}, \citenamefont {Cognet},\
  and\ \citenamefont {Pasquali}}]{Fakhri2010}%
  \BibitemOpen
  \bibfield  {author} {\bibinfo {author} {\bibfnamefont {N.}~\bibnamefont
  {Fakhri}}, \bibinfo {author} {\bibfnamefont {F.~C.}\ \bibnamefont
  {MacKintosh}}, \bibinfo {author} {\bibfnamefont {B.}~\bibnamefont {Lounis}},
  \bibinfo {author} {\bibfnamefont {L.}~\bibnamefont {Cognet}}, \ and\ \bibinfo
  {author} {\bibfnamefont {M.}~\bibnamefont {Pasquali}},\ }\href {\doibase
  10.1126/science.1197321} {\bibfield  {journal} {\bibinfo  {journal}
  {Science}\ }\textbf {\bibinfo {volume} {330}},\ \bibinfo {pages} {1804}
  (\bibinfo {year} {2010})}\BibitemShut {NoStop}%
\bibitem [{\citenamefont {Ward}\ \emph {et~al.}(2015)\citenamefont {Ward},
  \citenamefont {Hilitski}, \citenamefont {Schwenger}, \citenamefont {Welch},
  \citenamefont {Lau}, \citenamefont {Vitelli}, \citenamefont {Mahadevan},\
  and\ \citenamefont {Dogic}}]{Ward2015}%
  \BibitemOpen
  \bibfield  {author} {\bibinfo {author} {\bibfnamefont {A.}~\bibnamefont
  {Ward}}, \bibinfo {author} {\bibfnamefont {F.}~\bibnamefont {Hilitski}},
  \bibinfo {author} {\bibfnamefont {W.}~\bibnamefont {Schwenger}}, \bibinfo
  {author} {\bibfnamefont {D.}~\bibnamefont {Welch}}, \bibinfo {author}
  {\bibfnamefont {A.~W.~C.}\ \bibnamefont {Lau}}, \bibinfo {author}
  {\bibfnamefont {V.}~\bibnamefont {Vitelli}}, \bibinfo {author} {\bibfnamefont
  {L.}~\bibnamefont {Mahadevan}}, \ and\ \bibinfo {author} {\bibfnamefont
  {Z.}~\bibnamefont {Dogic}},\ }\href {\doibase 10.1038/NMAT4222} {\bibfield
  {journal} {\bibinfo  {journal} {Nature Materials}\ }\textbf {\bibinfo
  {volume} {14}},\ \bibinfo {pages} {583} (\bibinfo {year} {2015})}\BibitemShut
  {NoStop}%
\bibitem [{\citenamefont {Reisner}\ \emph {et~al.}(2005)\citenamefont
  {Reisner}, \citenamefont {Morton}, \citenamefont {Riehn}, \citenamefont
  {Wang}, \citenamefont {Yu}, \citenamefont {Rosen}, \citenamefont {Sturm},
  \citenamefont {Chou}, \citenamefont {Frey},\ and\ \citenamefont
  {Austin}}]{PhysRevLett.94.196101}%
  \BibitemOpen
  \bibfield  {author} {\bibinfo {author} {\bibfnamefont {W.}~\bibnamefont
  {Reisner}}, \bibinfo {author} {\bibfnamefont {K.~J.}\ \bibnamefont {Morton}},
  \bibinfo {author} {\bibfnamefont {R.}~\bibnamefont {Riehn}}, \bibinfo
  {author} {\bibfnamefont {Y.~M.}\ \bibnamefont {Wang}}, \bibinfo {author}
  {\bibfnamefont {Z.}~\bibnamefont {Yu}}, \bibinfo {author} {\bibfnamefont
  {M.}~\bibnamefont {Rosen}}, \bibinfo {author} {\bibfnamefont {J.~C.}\
  \bibnamefont {Sturm}}, \bibinfo {author} {\bibfnamefont {S.~Y.}\ \bibnamefont
  {Chou}}, \bibinfo {author} {\bibfnamefont {E.}~\bibnamefont {Frey}}, \ and\
  \bibinfo {author} {\bibfnamefont {R.~H.}\ \bibnamefont {Austin}},\ }\href
  {\doibase 10.1103/PhysRevLett.94.196101} {\bibfield  {journal} {\bibinfo
  {journal} {Phys. Rev. Lett.}\ }\textbf {\bibinfo {volume} {94}},\ \bibinfo
  {pages} {196101} (\bibinfo {year} {2005})}\BibitemShut {NoStop}%
\bibitem [{\citenamefont {Riefler}\ \emph {et~al.}(2010)\citenamefont
  {Riefler}, \citenamefont {Schmid}, \citenamefont {Burada},\ and\
  \citenamefont {H{\"a}nggi}}]{0953-8984-22-45-454109}%
  \BibitemOpen
  \bibfield  {author} {\bibinfo {author} {\bibfnamefont {W.}~\bibnamefont
  {Riefler}}, \bibinfo {author} {\bibfnamefont {G.}~\bibnamefont {Schmid}},
  \bibinfo {author} {\bibfnamefont {P.~S.}\ \bibnamefont {Burada}}, \ and\
  \bibinfo {author} {\bibfnamefont {P.}~\bibnamefont {H{\"a}nggi}},\ }\href
  {http://stacks.iop.org/0953-8984/22/i=45/a=454109} {\bibfield  {journal}
  {\bibinfo  {journal} {Journal of Physics: Condensed Matter}\ }\textbf
  {\bibinfo {volume} {22}},\ \bibinfo {pages} {454109} (\bibinfo {year}
  {2010})}\BibitemShut {NoStop}%
\bibitem [{\citenamefont {Bechinger}\ \emph {et~al.}(2016)\citenamefont
  {Bechinger}, \citenamefont {Di~Leonardo}, \citenamefont {L{\"o}wen},
  \citenamefont {Reichhardt}, \citenamefont {Volpe},\ and\ \citenamefont
  {Volpe}}]{Bechinger2016}%
  \BibitemOpen
  \bibfield  {author} {\bibinfo {author} {\bibfnamefont {C.}~\bibnamefont
  {Bechinger}}, \bibinfo {author} {\bibfnamefont {R.}~\bibnamefont
  {Di~Leonardo}}, \bibinfo {author} {\bibfnamefont {H.}~\bibnamefont
  {L{\"o}wen}}, \bibinfo {author} {\bibfnamefont {C.}~\bibnamefont
  {Reichhardt}}, \bibinfo {author} {\bibfnamefont {G.}~\bibnamefont {Volpe}}, \
  and\ \bibinfo {author} {\bibfnamefont {G.}~\bibnamefont {Volpe}},\
  }\href@noop {} {\bibfield  {journal} {\bibinfo  {journal} {Rev. Mod. Phys.}\
  }\textbf {\bibinfo {volume} {88}},\ \bibinfo {pages} {045006} (\bibinfo
  {year} {2016})}\BibitemShut {NoStop}%
\bibitem [{\citenamefont {Liu}\ \emph {et~al.}(2016)\citenamefont {Liu},
  \citenamefont {Zhou}, \citenamefont {Wang},\ and\ \citenamefont
  {Zhang}}]{liu2016bimetallic}%
  \BibitemOpen
  \bibfield  {author} {\bibinfo {author} {\bibfnamefont {C.}~\bibnamefont
  {Liu}}, \bibinfo {author} {\bibfnamefont {C.}~\bibnamefont {Zhou}}, \bibinfo
  {author} {\bibfnamefont {W.}~\bibnamefont {Wang}}, \ and\ \bibinfo {author}
  {\bibfnamefont {H.}~\bibnamefont {Zhang}},\ }\href@noop {} {\bibfield
  {journal} {\bibinfo  {journal} {Phys. Rev. Lett.}\ }\textbf {\bibinfo
  {volume} {117}},\ \bibinfo {pages} {198001} (\bibinfo {year}
  {2016})}\BibitemShut {NoStop}%
\bibitem [{\citenamefont {Wu}\ \emph {et~al.}(2015)\citenamefont {Wu},
  \citenamefont {Chen}, \citenamefont {Wang},\ and\ \citenamefont
  {Ai}}]{Wu2015a}%
  \BibitemOpen
  \bibfield  {author} {\bibinfo {author} {\bibfnamefont {J.~C.}\ \bibnamefont
  {Wu}}, \bibinfo {author} {\bibfnamefont {Q.}~\bibnamefont {Chen}}, \bibinfo
  {author} {\bibfnamefont {R.}~\bibnamefont {Wang}}, \ and\ \bibinfo {author}
  {\bibfnamefont {B.~Q.}\ \bibnamefont {Ai}},\ }\href {\doibase
  10.1063/1.4913491} {\bibfield  {journal} {\bibinfo  {journal} {Chaos}\
  }\textbf {\bibinfo {volume} {25}},\ \bibinfo {pages} {023114} (\bibinfo
  {year} {2015})}\BibitemShut {NoStop}%
\bibitem [{Sup()}]{SupportInfo}%
  \BibitemOpen
  \href@noop {} {\bibinfo  {journal} {See Supplemental Material for fabrication
  procedures, diffusion measurements, finite-element calculation, and Brownian
  dynamics simulations.}\ }\BibitemShut {NoStop}%
\bibitem [{\citenamefont {Happel}\ and\ \citenamefont
  {Brenner}(1965)}]{HappelBook}%
  \BibitemOpen
\bibfield  {journal} {  }\bibfield  {author} {\bibinfo {author} {\bibfnamefont
  {J.}~\bibnamefont {Happel}}\ and\ \bibinfo {author} {\bibfnamefont
  {H.}~\bibnamefont {Brenner}},\ }\href@noop {} {\emph {\bibinfo {title} {Low
  Reynolds Number Hydrodynamics}}}\ (\bibinfo  {publisher} {Prentice Hall,
  Englewood Cliffs, NJ},\ \bibinfo {year} {1965})\BibitemShut {NoStop}%
\bibitem [{\citenamefont {Tirado}\ \emph {et~al.}(1984)\citenamefont {Tirado},
  \citenamefont {Mart{\'\i}nez},\ and\ \citenamefont {de~la
  Torre}}]{Tirado1984}%
  \BibitemOpen
  \bibfield  {author} {\bibinfo {author} {\bibfnamefont {M.~M.}\ \bibnamefont
  {Tirado}}, \bibinfo {author} {\bibfnamefont {C.~L.}\ \bibnamefont
  {Mart{\'\i}nez}}, \ and\ \bibinfo {author} {\bibfnamefont {J.~G.}\
  \bibnamefont {de~la Torre}},\ }\href@noop {} {\bibfield  {journal} {\bibinfo
  {journal} {The Journal of chemical physics}\ }\textbf {\bibinfo {volume}
  {81}},\ \bibinfo {pages} {2047} (\bibinfo {year} {1984})}\BibitemShut
  {NoStop}%
\bibitem [{\citenamefont {Bitter}\ \emph {et~al.}(2017)\citenamefont {Bitter},
  \citenamefont {Yang}, \citenamefont {Duncan}, \citenamefont {Fairbrother},\
  and\ \citenamefont {Bevan}}]{Bitter2017}%
  \BibitemOpen
  \bibfield  {author} {\bibinfo {author} {\bibfnamefont {J.~L.}\ \bibnamefont
  {Bitter}}, \bibinfo {author} {\bibfnamefont {Y.}~\bibnamefont {Yang}},
  \bibinfo {author} {\bibfnamefont {G.}~\bibnamefont {Duncan}}, \bibinfo
  {author} {\bibfnamefont {H.}~\bibnamefont {Fairbrother}}, \ and\ \bibinfo
  {author} {\bibfnamefont {M.~A.}\ \bibnamefont {Bevan}},\ }\href {\doibase
  10.1021/acs.langmuir.7b01704} {\bibfield  {journal} {\bibinfo  {journal}
  {Langmuir}\ }\textbf {\bibinfo {volume} {33}},\ \bibinfo {pages} {9034}
  (\bibinfo {year} {2017})}\BibitemShut {NoStop}%
\bibitem [{\citenamefont {Lisicki}\ \emph {et~al.}(2016)\citenamefont
  {Lisicki}, \citenamefont {Cichocki},\ and\ \citenamefont
  {Wajnryb}}]{Lisicki2018}%
  \BibitemOpen
  \bibfield  {author} {\bibinfo {author} {\bibfnamefont {M.}~\bibnamefont
  {Lisicki}}, \bibinfo {author} {\bibfnamefont {B.}~\bibnamefont {Cichocki}}, \
  and\ \bibinfo {author} {\bibfnamefont {E.}~\bibnamefont {Wajnryb}},\ }\href
  {\doibase 10.1063/1.4958727} {\bibfield  {journal} {\bibinfo  {journal} {J.
  Chem. Phys.}\ }\textbf {\bibinfo {volume} {145}},\ \bibinfo {pages} {034904}
  (\bibinfo {year} {2016})}\BibitemShut {NoStop}%
\bibitem [{\citenamefont {Han}\ \emph {et~al.}(2009)\citenamefont {Han},
  \citenamefont {Alsayed}, \citenamefont {Nobili},\ and\ \citenamefont
  {Yodh}}]{Han2009}%
  \BibitemOpen
  \bibfield  {author} {\bibinfo {author} {\bibfnamefont {Y.}~\bibnamefont
  {Han}}, \bibinfo {author} {\bibfnamefont {A.}~\bibnamefont {Alsayed}},
  \bibinfo {author} {\bibfnamefont {M.}~\bibnamefont {Nobili}}, \ and\ \bibinfo
  {author} {\bibfnamefont {A.~G.}\ \bibnamefont {Yodh}},\ }\href {\doibase
  10.1103/PhysRevE.80.011403} {\bibfield  {journal} {\bibinfo  {journal} {Phys.
  Rev. E}\ }\textbf {\bibinfo {volume} {80}},\ \bibinfo {pages} {011403}
  (\bibinfo {year} {2009})}\BibitemShut {NoStop}%
\bibitem [{\citenamefont {Berezhkovskii}\ \emph {et~al.}(2015)\citenamefont
  {Berezhkovskii}, \citenamefont {Dagdug},\ and\ \citenamefont
  {Bezrukov}}]{Berezhkovskii2015}%
  \BibitemOpen
  \bibfield  {author} {\bibinfo {author} {\bibfnamefont {A.~M.}\ \bibnamefont
  {Berezhkovskii}}, \bibinfo {author} {\bibfnamefont {L.}~\bibnamefont
  {Dagdug}}, \ and\ \bibinfo {author} {\bibfnamefont {S.~M.}\ \bibnamefont
  {Bezrukov}},\ }\href {\doibase 10.1063/1.4934223} {\bibfield  {journal}
  {\bibinfo  {journal} {J. Chem. Phys.}\ }\textbf {\bibinfo {volume} {143}},\
  \bibinfo {pages} {164102} (\bibinfo {year} {2015})}\BibitemShut {NoStop}%
\bibitem [{\citenamefont {Goel}\ and\ \citenamefont
  {Richter-Dyn}(1974)}]{Goel1974}%
  \BibitemOpen
  \bibfield  {author} {\bibinfo {author} {\bibfnamefont {N.}~\bibnamefont
  {Goel}}\ and\ \bibinfo {author} {\bibfnamefont {N.}~\bibnamefont
  {Richter-Dyn}},\ }\href@noop {} {\emph {\bibinfo {title} {Stochastic Models
  in Biology}}}\ (\bibinfo  {publisher} {Academic Press},\ \bibinfo {year}
  {1974})\BibitemShut {NoStop}%
\end{thebibliography}
\end{document}